\documentclass[aps,amsfonts,amsmath,prd,preprint,nofootinbib]{revtex4}

\usepackage[utf8]{inputenc}
\usepackage{amsfonts}
\usepackage{physics}
\usepackage{mathtools}
\usepackage{mathrsfs}
\usepackage{amssymb}
\usepackage{appendix}
\usepackage{graphicx}
\usepackage{amsmath}
\usepackage{color}
\usepackage{feynmf}
\usepackage{simpler-wick} 

\usepackage[legalpaper, margin=1in]{geometry}
\begin{document}

\title{Bubble nucleation as seen by different observers}

\author{Yilin Chen and Alexander Vilenkin}
\address{Institute of Cosmology, Department of Physics and Astronomy, \\
Tufts University, Medford, Massachusetts 02155, USA}

\begin{abstract}

Pair production in a constant electric field is closely analogous to
bubble nucleation in a false vacuum.  The classical trajectories of the 
pairs are Lorentz invariant, but this invariance should be
broken by the nucleation process. Garriga {\it et al.} used a model detector, consisting of other particles 
interacting with the pairs, to investigate how pair production is seen by different Lorentzian observers.
{They found that particles (antiparticles) of the pair are predominantly observed moving in the direction of (opposite to) the electric field and} concluded that observers see pairs nucleating preferentially in the detector's rest frame.  Here, we apply this approach to the case where two detectors moving relative to one another are used to observe the particle and antiparticle of the same pair. We find that each detector will still observe nucleation to occur in its rest frame, regardless of the motion of the other detector.   However, if the relative velocity of the two detectors is sufficiently high, the particle and antiparticle of the pair can be observed moving towards one another with arbitrarily large momenta, contrary to the usual expectation.

\end{abstract}

\maketitle

\section{Introduction}

False vacuum decay by bubble nucleation is a fascinating process which raises some intriguing conceptual problems.  It was first discussed in the pioneering paper by Voloshin, Kobzarev and Okun \cite{Voloshin}, who argued that bubbles spontaneously nucleate at rest {and immediately start an accelerated expansion,} approaching the speed of light.  On the other hand, due to the Lorentz invariance of the false vacuum, bubbles do not have any preferred frame in which to nucleate.  The authors of \cite{Voloshin} suggested that the nucleation rate should include an integral over the Lorentz group, in order to account for all possible rest frames of nucleation.  This integral however is divergent and yields a meaningless infinite result for the nucleation rate.

An elegant resolution of this paradox was provided by Coleman \cite{Coleman}, who developed an instanton method for calculating the bubble nucleation rate.  The instanton solution in this case is $O(4)$ invariant, and its Lorentzian continuation, which describes time evolution of the bubble, is invariant with respect to Lorentz boosts.  
From this Coleman concluded that "an expanding bubble looks the same to all Lorentz observers, and to integrate over the Lorentz group is to erroneously count the same final state many times."

This, however, still leaves us with a puzzle.  The Lorentzian continuation of the instanton describes a bubble which contracts from infinite size, bounces at a minimum radius, and then re-expands.  It appears that the contracting part of the bubble worldsheet is unphysical and needs to be cut off.  But the cutoff would then break the Lorentz symmetry and define a preferred frame.

We can imagine an inertial observer using a set of detectors distributed in space, which are all at rest relative to the observer and which react to the presence of the bubble wall.  The question is then: What will be the nucleation surface in spacetime where the detectors will register the bubble to nucleate?


This issue was addressed by Garriga {\it et al.} in Refs.~\cite{Garriga1,Garriga2} using the close analogy between bubble nucleation and pair production in an electric field.  They considered a charged scalar field $\phi$ in a constant electric field $E$ in $(1+1)$ dimensions.  The field was assumed to be in the in-vacuum state, in which the initial hypersurface where the electric field is turned on and the quantum state of $\phi$ was prepared is removed to the infinite past, $t\to -\infty$.  It was shown in \cite{Garriga1} that this quantum state is Lorentz invariant.  
Particle-antiparticle pairs of the field $\phi$ are produced out of the vacuum at a constant rate \cite{Schwinger}.  In the semiclassical approximation, and with a suitable choice for the origin of time, the distance $2r$ between the particles of the pair satisfies
\begin{align}
r^2 -t^2 = r_0^2,
\label{worldsheet}
\end{align}
where the minimum radius is $r_0=m/eE$, $m$ is the particle's mass and $e$ is its charge.  The semiclassical approximation applies if $m^2/eE \gg 1$, when the nucleation rate of the pairs $\left(\propto \exp(-\pi m^2/eE)\right)$ is small.
The worldlines in Eq.~(\ref{worldsheet}) are Lorentz invariant, but it is usually assumed that the pair nucleates at minimal separation at $t=0$, so the parts of the worldlines where the particles move towards one another (corresponding to the contracting part of the bubble worldsheet) are cut off (see Fig.~1).  For a different inertial observer, whose time coordinate $t'$ is related to $t$ by a Lorentz transformation, the expected cutoff surface would be at $t'=0$, as illustrated in the right panel of Fig.~1.  From the point of view of the first observer this would correspond to a strange situation where the antiparticle is produced moving in the `wrong' direction and the particle appears at a later time.   

\begin{figure}
\begin{tabular}{cc}
     \includegraphics[width=0.45\textwidth]{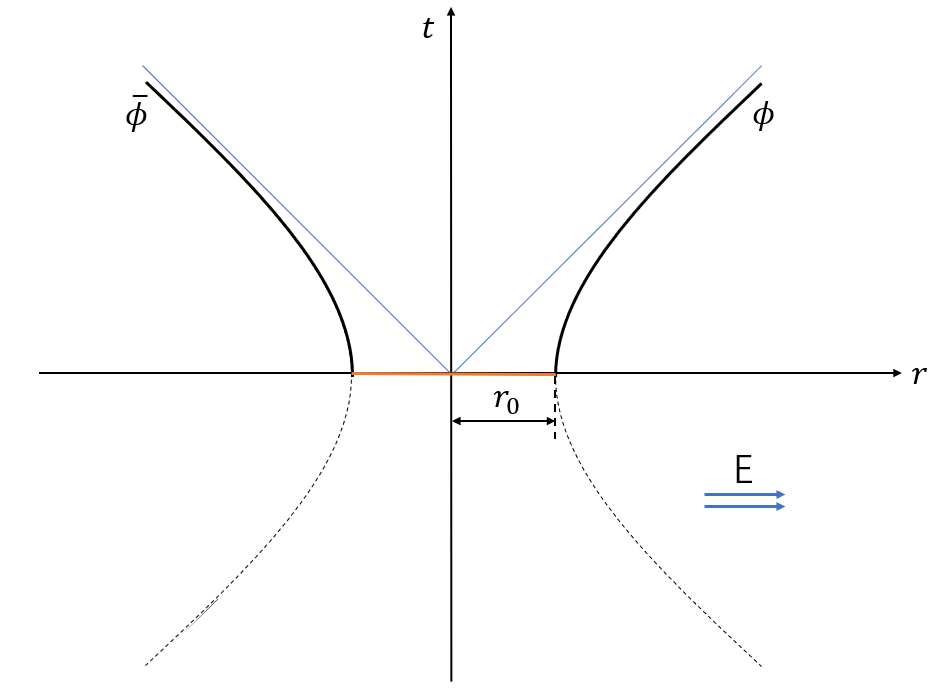}&\includegraphics[width=0.45\textwidth]{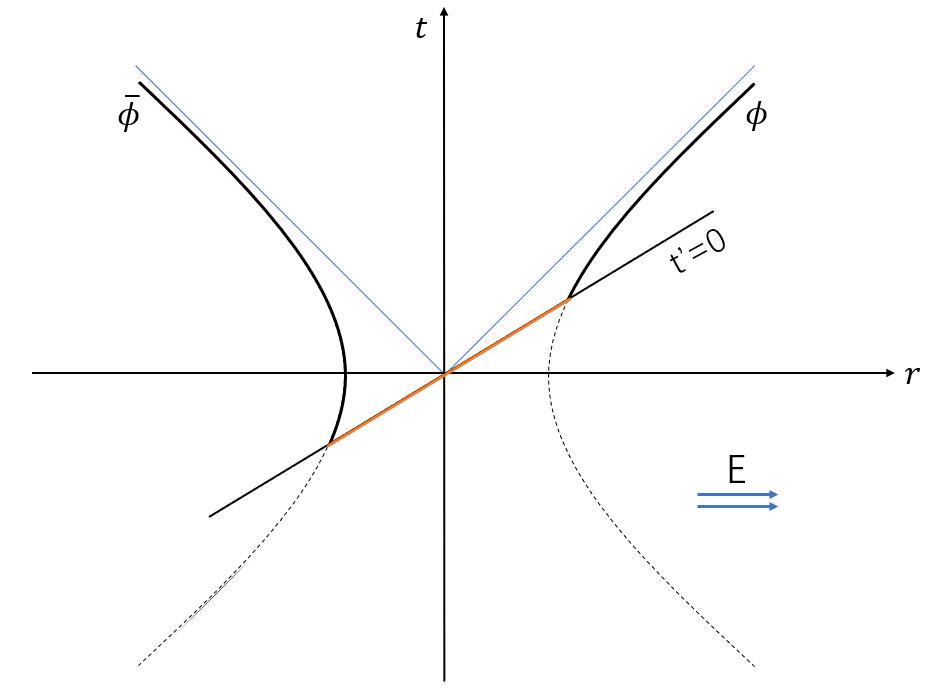}\\
     (a)&(b)\\
\end{tabular}
\caption{(a) The standard picture of pair nucleation.  $\phi$-particles and antiparticles nucleate at rest and are then driven apart by the electric field.  The solid black lines show the partcle's trajectories after nucleation; the dashed lines are the parts of classical trajectories that are usually assumed to be unphysical and are discarded.  (b) The same process viewed from a different reference frame, where the nucleation now occurs at the hypersurface $t'=0$.}
\end{figure}

In addition to $\phi$, Ref.~\cite{Garriga1} introduced another charged field $\psi$ and a real field $\chi$ which play the role of a detector.  The interaction between the fields was chosen of the form\footnote{This model was earlier studied by Massar and Parentani \cite{Massar} and by Gabriel {\it et al.} \cite{Gabriel} to investigate the Unruh effect for an accelerated detector.}
\begin{align}
{\cal H}_{int}=g(\phi\psi^\dagger \chi +{\rm h.c.}), 
\label{Lint}
\end{align}
where $g$ is a coupling constant.  If a $\psi$-particle is present in the initial state, it can annihilate a $\phi$-antiparticle via
$\psi{\bar\phi}\to\chi$.  One can then study the momentum distribution of $\chi$-particles in the final state to deduce the momentum distribution of the created pairs.  
The conclusion reached in Refs.~\cite{Garriga1,Garriga2} is that 
the pairs are predominantly observed to nucleate in the rest frame of the $\psi$-particle detectors.
This is consistent with the Lorentz invariance of the in-vacuum: in the setup considered in \cite{Garriga1,Garriga2}, the only frame that can determine the rest frame of nucleation is that of the $\psi$-detectors.  An observer comoving with the detectors will then see the nucleation picture illustrated in the left frame of Fig.~1, while other observers will see the distorted picture in the right frame of the figure.

It was noted in Ref.~\cite{Garriga1} that the in-vacuum state has some unphysical properties.  
An electric field which is turned on at $t\to -\infty$ would create an infinite density of pairs.  However, if the back-reaction of the pairs on the electric field is neglected, as it was done in \cite{Garriga1,Garriga2}, the dynamics of each particular pair is not affected by this infinite background.  One can then hope that the analysis based on the in-vacuum would agree with a more physical approach where the electric field is turned on for a finite time ${\cal T}$ in the limit when $\mathcal{T}$ gets very large.

The question we address in the present paper is what happens if two detectors moving relative to one another are used to observe the particle and antiparticle of the same pair.  As before, we assume the field $\phi$ to be in the in-vacuum state. We will interpret this state as approximating the state of $\phi$-particles in an electric field which is turned on for a long but finite time, so that the density of pairs is finite.
It will be convenient to use a slight generalization of the model (\ref{Lint}), which includes two sets of detector fields, with one set used to detect $\phi$-particles and the other to detect $\phi$-antiparticles.  Our goal will be to find out whether or not the two detectors can observe the "forbidden" part of history of the pair, where $\phi$ and ${\bar\phi}$ move towards one another.

The paper is organized as follows.  In Section II we introduce our detector model and the general formalism we use to calculate the distribution of $\chi$-particles in the final state.  This formalism is similar to, but somewhat different from that of Ref.~\cite{Garriga1}.  In Sec.~III we show how our formalism can be used to recover the results of \cite{Garriga1} for a single detector.
The $\chi$ particle distribution for the case of two detectors is calculated in Section IV, where we show that particles of the pair can in fact be observed moving towards one another. 
 Our conclusions are summarized and discussed in Section V.  Some technical details of the calculation are given in the Appendix.

\section{General formalism}

Particle-antiparticle pairs nucleating in a constant electric field $E$ in $(1+1)D$ are described by a complex field $\phi(x,t)$,
\begin{align}
  \phi\!&=\!\int\frac{dk}{\sqrt{2\pi}}\left(a_k^{in} \phi_k^{in}+b^{in \dagger}_{-k} \phi^{in *}_k\right)e^{ikx},\  \phi^{\dagger}\!=\!\int\frac{dk}{\sqrt{2\pi}}\left(b_{-k}^{in}\phi_k^{in}+a^{in \dagger}_{k}\phi^{in *}_k\right)e^{-ikx},
\end{align}  
where $a_k^{in}$ and $b_k^{in}$ are respectively the particle and antiparticle annihilation operators in the in-vacuum state $\ket{0,in}$,
\begin{align}
a_k^{in}\ket{0,in}=b_k^{in}\ket{0,in}=0.
\end{align}
We also introduce four additional fields, $\psi_j$ and $\chi_j$, $j=1,2$, which will play the role of detectors.  $\psi_j$ are complex fields, 
\begin{align}
    \psi_j \!&=\!\int\frac{dq}{\sqrt{2\pi}}\left(d_{q(j)}^{in}\psi_{q}^{in}
    +f^{in \dagger}_{-q(j)}\psi^{in *}_{q}\right)e^{iqx},\ \psi_j^{\dagger}\!=\!\int\frac{dq}{\sqrt{2\pi}}\left(f_{-q(j)}^{in}\psi_{q}^{in}+d^{in \dagger}_{q(j)}\psi^{in *}_{q}\right)e^{-iqx},
\end{align}
and $\chi_j$ are real,
\begin{align}
    \chi_j\!&=\!\int\frac{dp}{\sqrt{2\pi}}\left(c_{p(j)} \chi_{p}+c^{\dagger}_{-p(j)}\chi^*_{p}\right)e^{ipx}.
\end{align}
We assume for simplicity that the fields $\psi_j$ have the same mass $m_\psi$ and the field $\chi_j$ have mass $m_\chi$.  In what follows, we shall always use the notation $k$, $q$ and $p$ to denote the momenta of $\phi$, $\psi_j$ and $\chi_j$ particles, respectively.  

We shall assume that the model satisfies the mass hierarchy: 
\begin{align}
m_{\chi}\gg m_{\psi} \gg m_{\phi}, 
\label{hierarchy}
\end{align}
or equivalently
\begin{align}
\lambda_\chi \gg \lambda_\psi \gg \lambda_\phi \gg 1,
\label{lambdahierarchy}
\end{align}
where $\lambda=m^2/eE$ for each kind of particles and
the last inequality is needed for pair nucleation rate to be exponentially suppressed, so that pairs can be described semiclassically.  The second inequality in (\ref{hierarchy}) ensures that pair production of the detector $\psi$-particles can be neglected (compared to that of $\phi$-particles).  The first inequality in (\ref{hierarchy}) can be relaxed to $m_\chi \gtrsim m_\psi$, as it was done in Ref.~\cite{Garriga2}.  But assuming the strong inequality simplifies the calculation and will be sufficient for our purposes here.  

The in-vacuum mode functions of the field $\phi$ can be expressed in terms of the parabolic cylinder functions,
\begin{align}
\phi_k^{in}(t)=\frac{1}{(2eE)^{1/4}}e^{i\frac{\pi}{4}\nu^*}D_{\nu^*}\left[-\sqrt{2}e^{-i\pi/4} z\right],
\label{phik}
\end{align}
where 
\begin{align}
z=\sqrt{eE}(t+k/eE), ~~~~ \nu=-\frac{1+i\lambda_\phi}{2}, ~~~~ \lambda_\phi=\frac{m_\phi^2}{eE}.
\end{align}
Here, $k$ is the canonical momentum of $\phi$-particles, while the physical momentum changes with time due to the acceleration by the electric field,
\begin{align}
k_{phys}=k+eEt,
\label{kphys}
\end{align}
and opposite sign for antiparticles.

The in-vacuum mode functions $\psi_q$ for the fields $\psi_j$ are given by Eq.~(\ref{phik}) with $k$ replaced by $q$ and $\lambda_\phi$ replaced by $\lambda_\psi=m_\psi^2/eE$, and the mode functions for $\chi_j$ are
\begin{align}
\chi_p(t) = (2\omega_p)^{-1/2}e^{-i\omega_p t},
\end{align}
where $\omega_p=(p^2+m_\chi^2)^{1/2}$.  Similarly to (\ref{kphys}), the physical momentum of $\psi$-particles is
\begin{align}
q_{phys}=q+eEt.
\end{align}

The fields $\phi$ and $\psi$ can also be expanded in the basis of out-vacuum mode functions, e.g.,
\begin{align}
  \phi\!&=\!\int\frac{dk}{\sqrt{2\pi}}\left(a_k^{out} \phi_k^{out}+b^{out \dagger}_{-k} \phi^{out *}_k\right)e^{ikx},\  \phi^{\dagger}\!=\!\int\frac{dk}{\sqrt{2\pi}}\left(b_{-k}^{out}\phi_k^{out}+a^{out \dagger}_{k}\phi^{out *}_k\right)e^{-ikx},
\end{align}  
where the out-vacuum annihilation operators satisfy
\begin{align}
a_k^{out}\ket{0,out}=b_k^{out}\ket{0,out}=0
\end{align}
and the mode functions are given by
\begin{align}
\phi_k^{out}(t)=\frac{1}{(2eE)^{1/4}}e^{-i\frac{\pi}{4}\nu}D_{\nu}\left[\sqrt{2}e^{i\pi/4} z\right] =\phi_{-k}^{in *}(-t).
\label{phikout}
\end{align}
The in and out mode functions are related by the Bogoliubov transformation
\begin{align}
\phi_k^{in}=\alpha_\phi \phi_k^{out}+\beta_\phi \phi_k^{out *},
\end{align}
where
\begin{align}
\alpha_\phi=\frac{\sqrt{2\pi}}{\Gamma(-\nu^*)}e^{i\frac{\pi}{4}(\nu^*-\nu)}, ~~~~ \beta_\phi=e^{i\pi\nu^*}.
\end{align}
These transformation coefficients satisfy
\begin{align}
{\abs{\beta_\phi}}^2=e^{-\pi\lambda_\phi}=\exp\left(-\frac{\pi m_\phi^2}{eE}\right),
\end{align}
\begin{align}
{\abs{\alpha_\phi}}^2=1+{\abs{\beta_\phi}}^2.
\end{align}
The quantity ${\abs{\beta_\phi}}^2\ll 1$ is proportional to the $\phi$-pair creation rate.

The in-vacuum can also be expressed in terms of the out-vacuum as (see, e.g., \cite{Gabriel})
\begin{align}
\ket{0,in}_\phi = Z_\phi \exp\left[\frac{\beta_\phi^*}{\alpha_\phi^*}\int dk a_k^{out \dagger}b_{-k}^{out \dagger} \right] \ket{out,0}_\phi = Z_\phi \sum_n \frac{1}{n\!} \left(\frac{\beta_\phi^*}{\alpha_\phi^*}\right)^n \left(\int dk a_k^{out \dagger}b_{-k}^{out \dagger} \right)^n \ket{0,out}_\phi
\label{0in}
\end{align}
Here,
\begin{align}
Z_\phi=~_\phi\bra{0,out}\ket{0,in}_\phi
\end{align}
is the amplitude of producing no $\phi$-pairs and the $n$-th term in the Taylor expansion corresponds to a state with $n$ pairs at late times.
Similar relations can be written for the $\psi$ field.  The in-vacuum of all fields is just the tensor product 
\begin{align}
\ket{0,in}=\ket{0,in}_\phi \ket{0,in}_{\psi (1)} \ket{0,in}_{\psi (2)} \ket{0}_{\chi(1)} \ket{0}_{\chi(2)}
\end{align}
and similarly for out-vacuum.  Since there is no pair production of $\chi$ particles, the in- and out-vacua for the $\chi$ fields are the same.

{The amplitude $Z_\phi$ for producing no $\phi$-pairs vanishes in an infinite spacetime volume.  We shall therefore assume implicitly that the electric field is turned on for a finite time ${\cal T}$, with the limits ${\cal T}\to\infty$ and $E\to 0$ taken at the end of the calculation.  The limits are assumed to be taken in such a way that the probability of producing a pair per unit length ($\sim |\beta_\phi|^2{\cal T}$) remains finite.  As we already mentioned, due to the mass hierarchy (\ref{hierarchy}) the rate of creation of $\psi$-pairs is negligible compared to that of $\phi$-pairs.  Hence we will set $Z_\psi=1$.}

The interaction Hamiltonian density for our model is
\begin{align}
{\cal H}_{int} = g_1 \phi^\dagger \psi_1 \chi_1  +  g_2 \phi^\dagger \psi_2 \chi_2  +  h.c .
\label{Hint}
\end{align}
We are interested in the process where $\psi_1$ and ${\bar\psi}_2$ particles in the initial state collide respectively with $\phi$-antiparticle and particle of the same pair, producing $\chi_1$ and $\chi_2$ particles in the final state, as illustrated in Fig.~2.  Hence we choose the initial state as
\begin{align}
    \ket{i}=d^{\dagger}_{q(1)} f^{\dagger}_{q'(2)}\ket{0,in}\equiv\ket{q \bar{q}'},
\label{i}
\end{align}
where the bar over $q'$ indicates that this is the momentum of an antiparticle, and to streamline the notation in the last step we suppressed the indices (1) and (2) specifying the type of $\psi$-particles. 

\begin{figure}
    \centering
    \includegraphics{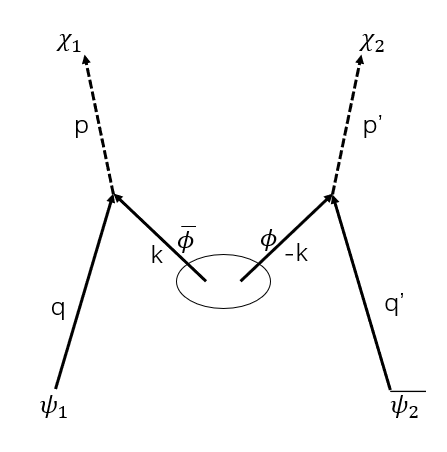}
    \caption{$\psi_1$ and $\bar{\psi_2}$ detectors collide respectively with $\bar{\phi}$ and $\phi$ particles of a $\phi\bar{\phi}$ pair, producing $\chi_1$ and $\chi_2$ particles in the final state.}
\end{figure}

The state (\ref{i}) evolves into
\begin{align}
    \ket{f}=S\ket{i},
\end{align}
at $t\to\infty$, where
\begin{align}\label{sm}
    S=1-i\int_{-\infty}^{\infty}dt H_{int}(t)-\frac{1}{2}\int_{-\infty}^{\infty}dt\int_{-\infty}^{\infty}dt' T\{H_{int}(t) H_{int}(t')\}+...\equiv1+S^{(1)}+S^{(2)}+...
\end{align}
is the $S$-matrix.  The $\chi_1$-particle distribution in the final state is given by
\begin{align}\label{Nchi}
    \frac{dN_{\chi}}{dp}=\frac{\bra{f}c^{\dagger}_{p(1)}c_{p(1)}\ket{f}}{\braket{i}{i}}.
\end{align}

For the process $\psi_1 {\bar\psi}_2 \to \chi_1 \chi_2$ that we are interested in,
the lowest order in $g_i$ contribution to \eqref{Nchi} comes from the second order term in \eqref{sm},
\begin{align}\label{bra}
    \bra{f}c^{\dagger}_pc_p\ket{f}\to\int dp_1\int dp'_1\int dp_2\int dp'_2 \bra{q\bar{q}'}S^{\dagger}_2\ket{p_1p'_1}\bra{p_1p'_1}c^{\dagger}_pc_p\ket{p_2p'_2}\bra{p_2p'_2}S_2\ket{q\bar{q}'},
\end{align}
where we have inserted two sets of intermediate states defined in the out-vacuum,
\begin{align}
\ket{p p'}\equiv c_{p(1)}^\dagger c_{p'(2)}^\dagger \ket{0,out}.
\end{align}
Now, using the relation
\begin{align}
    \bra{p_1p'_1}c^{\dagger}_{p(1)}c_{p(1)}\ket{p_2p'_2}=\delta(p_1-p)\delta(p-p_2)\delta(p'_1-p'_2)
\end{align}
we find
\begin{align}\label{cM}
    \bra{f}c^{\dagger}_{p(1)} c_{p(1)} \ket{f}=\int dp' \abs{\mathcal{M}(p,p'|q,\bar{q}')}^2,
\end{align}
where
\begin{align}
    \mathcal{M}=\int dt\int dt' \bra{pp'}T\{H_{int}(t)H_{int}(t')\}\ket{q\bar{q}'}
\label{mathcalM}
\end{align}
has the meaning of the scattering amplitude for the process $\ket{q{\bar q}'}\to \ket{pp'}$.

\section{Single detector}

\subsection{Constant coupling}

We shall first show how the results of Ref.~\cite{Garriga1} can be reproduced in our formalism in the case of a single $\psi$ particle detector.  In this case the initial state is
\begin{align}
    \ket{i}=d^{\dagger}_{q} \ket{0,in}\equiv\ket{q},
\label{isingle}
\end{align}
the scattering process is $\psi \to \chi\phi$, and the lowest order in $g$ contribution to \eqref{Nchi} comes from the first order term in \eqref{sm}:
\begin{align}\label{bra2}
    \bra{f}c^{\dagger}_pc_p\ket{f}\to\int dp_1\int dk_1\int dp_2\int dk_2 \bra{q}S^{\dagger}_1\ket{p_1k_1}\bra{p_1k_1}c^{\dagger}_pc_p\ket{p_2k_2}\bra{p_2k_2}S_1\ket{q},
\end{align}
where we have inserted two sets of intermediate states
\begin{align}
\ket{p k}\equiv c_{p}^\dagger {a_{k}^{out}}^\dagger \ket{0,out}.
\end{align}
In this section we do not distinguish between the two kinds of $\psi$ and $\chi$ particles.  Using that
\begin{align}
    \bra{p_1k_1}c^{\dagger}_{p}c_{p}\ket{p_2k_2}=\delta(p_1-p)\delta(p-p_2)\delta(k_1-k_2)
\end{align}
we obtain
\begin{align}\label{cMsingle}
    \bra{f}c^{\dagger}_{p} c_{p} \ket{f}=\int dk \abs{\mathcal{M}(p,k|q)}^2,
\end{align}
where
\begin{align}
    \mathcal{M}(p,k;q)=\int dt\bra{pk} H_{int}(t)\ket{q}.
\end{align}

To calculate the amplitude $\mathcal{M}$, we shall use the representation (\ref{0in}) for the in-vacuum in the initial state (\ref{isingle}).  Only the $n=1$ term gives a nonzero contribution.  It corresponds to the process where $\psi$ scatters on the $\phi$ antiparticle of a pair, while the $\phi$ particle propagates to the final state:
\begin{align}
\mathcal{M}=g Z_\phi \frac{\beta^*}{\alpha^*}\int dx \int dt \int dk' \bra{0,out} c_{p} a_{k} \{\phi^\dagger\psi \chi\}(x,t) d_q^\dagger  a_{k'}^{out \dagger} b_{-k'}^{out \dagger} \ket{0,out}.
\end{align}
A straightforward calculation gives
\begin{align}
\mathcal{M}=gZ_\phi \frac{\beta_\phi^*}{\alpha_\phi^*}\delta(p+k-q) \mathcal{A},
\label{MA}
\end{align}
where
\begin{align}
\mathcal{A}=\int dt \psi_q^{out}(t)\phi_{q-p}^{out}(t)\chi_{-p}^*(t).
\label{Asingle}
\end{align}
Here we have used an expansion of the field operators $\phi$ and $\psi$ in out-vacuum modes. 

To make a connection with the calculations in Ref.~\cite{Garriga1}, it will be useful to express the amplitude $\mathcal{A}$ in terms of in-vacuum mode functions.  To this end, we use Eq.~(\ref{phikout}) and change the integration variable $t$ in (\ref{Asingle}) to $-t$.  This gives
\begin{align}
\mathcal{A}= \int dt \psi_{-q}^{in *}(t)\phi_{p-q}^{in *}(t) \chi_{-p}(t) ,
\label{Ai}
\end{align}
where we have used that $\chi_{-p}^*(-t)=\chi_{-p}(t)$.  

The integral in Eq.~(\ref{Ai}) was calculated in Ref.~\cite{Garriga1}, where it was denoted $\mathcal{A}_\psi (-q,-p)$. 
With account taken of the mass hierarchy (\ref{hierarchy}), it can be simplified to (see Appendix A)
\begin{align}
\abs{\mathcal{A}}\approx  \frac{1}{\sqrt{2eE} m_\chi}\sqrt{\frac{\pi}{\omega_p}}.
\label{Aapprox}
\end{align}
Substituting this in Eq.~(\ref{MA}) for $\mathcal{M}$ and then in Eq.~(\ref{cMsingle}), we find
\begin{align}
\label{cMcm}
    \bra{f}c^{\dagger}_{p} c_{p} \ket{f}=\delta(0) {\abs{Z _\phi}}^2{\abs{\frac{\beta_\phi}{\alpha_\phi}}}^2 \frac{\pi g^2}{2eE m_\chi^2 \omega_p}.
\end{align}

The infinite factor $\delta(0)$ appears in the above equation because we used plane wave states normalized to a $\delta$-function,
\begin{align}
\bra{q}\ket{q'}=\delta(q-q').
\end{align}
Then the denominator of Eq.~(\ref{Nchi}) is $\bra{q}\ket{q}=\delta(0)$ and the momentum distribution of $\chi$-particles becomes
\begin{align}\label{Nchipsingle}
    {dN_{\chi}}= g^2 {\abs{Z _\phi}}^2{\abs{\frac{\beta_\phi}{\alpha_\phi}}}^2 {\abs{\mathcal{A}}}^2 dp 
= {\abs{Z _\phi}}^2{\abs{\frac{\beta_\phi}{\alpha_\phi}}}^2\frac{\pi^2 g^2} {2eEm^2_{\chi}}\frac{dp}{\omega_{p}}.
\end{align}
Apart from the factor ${\abs{Z _\phi}}^2$, this is in agreement with the result of \cite{Garriga1}.

\subsection{Time-dependent coupling}

The $\chi$-particle distribution in Eq.~(\ref{Nchipsingle}) is Lorentz invariant, so it cannot be used to determine the rest frame of pair nucleation.  To address this problem, Ref.~\cite{Garriga1} introduced a variable coupling, such that the detector is turned on only for a finite period of time $\Delta t\sim T$ around $t=0$ in the rest frame of the detector.  This was implemented by replacing the coupling constant $g$ with a variable coupling $g(t)=g f(t)$, where
\begin{align}
f(t)=\exp(-t^2/T^2).
\label{ft}
\end{align}													
The canonical momentum of $\psi$ was chosen as $q=0$, so that the physical momentum at $t=0$ is $q_{phys}(0)=0$.  Furthermore, it was assumed that 
\begin{align}
m_\psi \gg eET \gg m_\phi ,
\label{eET}
\end{align}
The first inequality in (\ref{eET}) ensures that $\psi$ detectors are not significantly accelerated by the electric field during the time $T$ when the detector is on. 
We note that it follows from Eq.~(\ref{eET}) that
\begin{align}
eET^2 \gg \frac{m_\phi^2}{eE} =\lambda_\phi \gg 1.
\label{eET2}
\end{align}

For any physical momentum $q_{phys}$ of the $\psi$ detector, the classical kinematics of the scattering process $\psi{\bar\phi}\to\chi$ allows only two values for the momentum $k_{phys}$ of ${\bar\phi}$: one where ${\bar\phi}$ hits $\psi$ from the left and the other where it hits from the right \cite{Garriga1}:  
\begin{align}
2m_\psi^2 k_{phys}=M^2 q_{phys}\pm \sqrt{q_{phys}^2+m_\psi^2} \sqrt{M^4-4m_\phi^2 m_\psi^2},
\label{kq}
\end{align}
where $M^2\equiv m_\chi^2 -m_\psi^2 -m_\phi^2$.  
The upper and lower signs in Eq.~(\ref{kq}) correspond to ${\bar\phi}$ hitting from the left and from the right, respectively.\footnote{Note that our notation differs from that in \cite{Garriga1}, where they denote the momentum of ${\bar\phi}$ by $-k$.}   With the electric field directed to the right $(E>0)$, it was shown in Ref.~\cite{Garriga1,Garriga2} that ${\bar\phi}$-particles hit predominantly from the right, that is with $k_{phys}<0$.  Collisions with $k_{phys}>0$ are exponentially suppressed.  


Since our goal is to study two detectors moving relative to one another, we will need to deal with a more general situation, when the momentum of the detector particle is $q\neq 0$. It will be sufficient for our purposes, and will also simplify the analysis, to consider ultrarelativistic detectors with $|q|\gg m_\psi$. The classical equation (\ref{kq}) holds only approximately, but it proved to be fairly accurate in Refs.~\cite{Garriga1,Garriga2}, so we will use it as a guide for what one can expect in a quantum scattering process.
Expanding Eq.~(\ref{kq}) in small parameters $m_\psi/|q|$ and $m_\phi m_\psi/M^2$, we obtain
\begin{align}
2m_\psi^2 k_{phys}=M^2\left[q-|q|-\frac{m_\psi^2}{2|q|}+\frac{2m_\phi^2 m_\psi^2}{M^4}|q|\right].
\label{kq2}
\end{align}
Here we used the lower sign in (\ref{kq}), anticipating that ${\bar\phi}$ particles will hit predominantly from the right.  With a further assumption that
\begin{align}
|q|\gg M^2/m_\phi
\end{align}
we can neglect the third term in square brackets of Eq.~(\ref{kq2}).  Then we find
\begin{align}
k_{phys}\approx p \approx \left(\frac{M}{m_\psi}\right)^2 q ~~{\rm for}~~ q<0
\label{kpq1}
\end{align}
and
\begin{align}
k_{phys}\approx  \left(\frac{m_\phi}{M}\right)^2 q,~~~p\approx q ~~{\rm for}~~ q>0,
\label{kpq2}
\end{align}
where $p=q_{phys}+k_{phys}$ is the momentum of the $\chi$-particle.
This suggests that for sufficiently large positive $q$ the ${\bar\phi}$ particle can be detected moving with a large momentum in the "wrong" direction (that is, to the right).  

Let us now turn to a quantum description of the scattering.
For a time-dependent coupling Eq.~(\ref{Ai}) is replaced by
\begin{align}
\mathcal{A}= \int dt f(t) \psi_{-q}^{in *}(t)\phi_{p-q}^{in *}(t) \chi_{-p}(t) .
\label{A-p-q}
\end{align}
Note that since $f(-t)=f(t)$, we could still use the change of integration variable $t\to -t$ to transform from out- to in-modes, as we did in Eqs.~(\ref{Asingle}),(\ref{Ai}).
The $\psi$ mode function appearing in Eq.~(\ref{A-p-q}) is given by
\begin{align}
\psi_{-q}^{in*}(t)=\frac{1}{(2eE)^{1/4}}e^{-i\frac{\pi}{4}\nu_\psi}D_{\nu_\psi}\left[-\sqrt{2}e^{i\pi/4} z\right],
\label{psiq}
\end{align}
where $z=\sqrt{eE}(t-q/eE)$.  With $f(t)$ from Eq.~(\ref{ft}), we shall assume that 
\begin{align}
|q|\gg eET,
\label{qeET}
\end{align}
so the physical momentum $q_{phys}$ does not change significantly during the time when the detector is on.  
Imposing a further constraint on $q$,
\begin{align}
|q|\gg \frac{m_\psi^2}{\sqrt{eE}},
\label{condition2}
\end{align}
we can use the asymptotic form of the parabolic cylinder function in (\ref{psiq}) with $|z|\gg |\nu_\psi|$.  For $q>0$ and $|t|\lesssim T$ we have $z<0$ and \cite{GR}
\begin{align}
D_{\nu_\psi}\left[-\sqrt{2}e^{i\pi/4} z\right]\approx  2^{\nu_\psi/2}e^{i\pi\nu_\psi/4}(-z)^{\nu_\psi} e^{-i\frac{z^2}{2}} 
\label{Dnupsi}
\end{align}
The factor $(-z)^{\nu_\psi}$ can be represented as
\begin{align}
(-z)^{\nu_\psi}=\left(\frac{q}{\sqrt{eE}}\right)^{-\frac{1}{2}(1+i\lambda_\psi)} e^{\nu_\psi \ln\left(1-\frac{eEt}{q}\right)} \approx \left(\frac{\sqrt{eE}}{q}\right)^{1/2} e^{-\nu_\psi \frac{eEt}{q}},
\label{-z}
\end{align}
where in the last step we have expanded the logarithm in Taylor series, keeping only the first term.  This term can be large for some parameter values, while it can be verified that the next term is $\ll 1$.  We have also dropped an irrelevant overall phase factor. (We will continue dropping such factors in what follows.) 

Combining Eqs.~(\ref{psiq}), (\ref{Dnupsi}) and (\ref{-z}), we can write
\begin{align}
\psi_{-q}^{in*}(t) \approx \frac{1}{\sqrt{2q}} e^{iqt}e^{i(m_\psi^2/2q)t}e^{-\frac{i}{2}eEt^2} \approx \frac{1}{\sqrt{2q}} e^{i\epsilon_q t} e^{-\frac{i}{2}eEt^2}~~~~(q>0)
\label{psiq1}
\end{align}
where
\begin{align}
\epsilon_q=\sqrt{q^2+m_\psi^2}\approx q\left(1+\frac{m_\psi^2}{2q^2}\right).
\end{align}
Similarly, for $q<0$, $z>0$ we obtain
\begin{align}
\psi_{-q}^{in*}(t) \approx \frac{1}{\sqrt{-2q}} e^{i\epsilon_q t} e^{\frac{i}{2}eEt^2}~~~~(q<0)
\label{psiq2}
\end{align}

We will now impose additional restrictions on $q$ which will allow us to use the approximate forms (\ref{psiq1}) and (\ref{psiq2}) for the $\phi$ mode function in Eq.~(\ref{A-p-q}).  For $q>0$ we expect the amplitude to be peaked at 
\begin{align}
k=p-q\approx \left(\frac{m_\phi}{M}\right)^2 q.
\end{align}
{The conditions analogous to Eqs.~(\ref{qeET}), (\ref{condition2}) are $|k|\gg eET$ and 
$z_\phi\approx (p-q)/{\sqrt{eE}} \gg |\nu_\phi|\approx m_\phi^2/2eE$.   They imply the following restrictions on $q$:
\begin{align}
|q|\gg \frac{m_\chi^2}{m_\phi^2}eET,
\label{condition4}
\end{align}
\begin{align}
q\gg \frac{M^2}{\sqrt{eE}}\approx \frac{m_\chi^2}{\sqrt{eE}}.
\label{condition3}
\end{align}
Note that with the conditions (\ref{condition4}), (\ref{condition3}) the constraints (\ref{qeET}) and (\ref{condition2}) are also satisfied.}

Now, if $q$ obeys the conditions (\ref{condition4}), (\ref{condition3}), we can estimate the amplitude (\ref{A-p-q}) using the asymptotic forms of the mode functions $\psi_{-q}^*$ and $\phi_{p-q}^*$:
\begin{align}
\mathcal{A}\approx \frac{1}{\sqrt{8q(p-q)\omega_p}} \int dt \exp\left[-\frac{t^2}{T^2}+i(\epsilon_q+{\xi}_{p-q} -\omega_p)t\right] 
\end{align}
\begin{align}
= \frac{\sqrt{\pi} T}{\sqrt{8q(p-q)\omega_p}}\exp\left[-\frac{T^2}{4}(\epsilon_q+{\xi}_{p-q} -\omega_p)^2\right],
\label{A2}
\end{align}
where $\xi_k=\sqrt{k^2+m_\phi^2}$ and we have again dropped an overall phase factor.  Note that the factors $\exp\left(\pm \frac{i}{2}eEt^2\right)$ cancelled out in the integrand, because we used the asymptotic form of $\psi_{-q}^*$ for $-q<0$ and that of $\phi_{p-q}^*$ for $p-q>0$.  

The exponential factor in Eq.~(\ref{A2}) can be interpreted as expressing energy conservation in the collision with an accuracy $\sim 2/T$.
As expected, the amplitude (\ref{A2}) has a peak at the values (\ref{kpq2}), where it can be approximately expressed as
\begin{align}
\mathcal{A}\approx \sqrt{\frac{\pi}{2}}\frac{m_\chi T}{2m_\phi q^{3/2}}\exp\left[-\frac{m_\chi^8 T^2}{16m_\phi^4 q^4} \left(p-q-\frac{m_\phi^2}{m_\chi^2}q \right)^2\right].
\label{Aapprox3}
\end{align}
The top of the peak is at $k=p-q= \left(\frac{m_\phi}{m_\chi}\right)^2 q$ and its width is
\begin{align}
\frac{\delta k}{k}\sim \frac{q}{m_\chi^2 T}.
\end{align} 
{We assumed that $q$ satisfies the conditions (\ref{condition4}), (\ref{condition3}), which imply respectively that $\delta k/k\gg eE/m_\phi^2$ and $\delta k/k \gg 1/\sqrt{eE}T$.  In both cases the right-hand side is $\ll 1$, so we can choose the parameters so that $\delta k/k\ll 1$ and
the momentum of ${\bar\phi}$ can be measured with a good accuracy.}
 
For $q$ large and negative we have $-q>0$ and $p-q<0$, so again the factors $\exp\left(\pm \frac{i}{2}eEt^2\right)$ cancel out in the time integral and the amplitude is still given by Eq.~(\ref{Aapprox3}).  In this case the peak is located at the values (\ref{kpq1}) and
\begin{align}
\mathcal{A}\approx \sqrt{\frac{\pi}{2}}\frac{m_\psi T}{2m_\chi |q|^{3/2}}\exp\left[-\frac{m_\psi^8 T^2}{16 m_\chi^4 q^4} \left(p-\frac{m_\chi^2}{m_\psi^2}q \right)^2\right].
\label{Aapprox2}
\end{align}
The conditions on $q$ in this case are less restrictive.

We have thus found that a $\psi$ detector moving with a large positive momentum can detect a ${\bar\phi}$ particle moving to the right, in agreement with classical expectations.  Note that this is not in contradiction with the results of Ref.~\cite{Garriga1}: the ${\bar\phi}$ particle is moving to the left in the rest frame of the detector. Eqs.~(\ref{kpq1}), (\ref{kpq2}) can of course be obtained using a Lorentz transformation from the frame where $q=0$.  An interesting situation arises when we have two detectors moving towards one another with large Lorentz factors.  This will be discussed in the next Section.

\section{Two detectors}

Let us now go back to the scattering amplitude (\ref{mathcalM}) for two detectors.  
Once again we use the representation (\ref{0in}) for the in-vacuum in the initial state (\ref{i}).  Only $n=0$ and $n=1$ terms give nonzero contributions,
\begin{align}
\mathcal{M}=\mathcal{M}^{(0)} +\mathcal{M}^{(1)}.
\end{align}
$\mathcal{M}^{(1)}$ corresponds to the process we are interested in (illustrated in Fig.~2), where ${\bar\psi}_2$ and $\psi_1$ scatter on the $\phi$ particle and antiparticle of the same pair, while $\mathcal{M}^{(0)}$ corresponds to a direct scattering $\psi_1 {\bar\psi}_2 \to \chi_1 \chi_2$ by exchange of a virtual $\phi$ particle.  
{The Feynman diagram for the latter process is shown in Fig.~3.  The two processes can be easily distinguished observationally (in the case of a time-dependent coupling).  For a direct scattering, the total energy of the $\chi$ particles in the final state is approximately equal to that of the initial $\psi$ particles.  (This is because the energy is approximately conserved during the collision.)   On the other hand, for the process shown in Fig.~2 the final energy can be significantly larger, due to the contribution of the $\phi{\bar\phi}$ pair.  We will focus on the calculation of $\mathcal{M}^{(1)}$ in what follows.}

\begin{figure}
    \centering
    \includegraphics[width=0.65\textwidth]{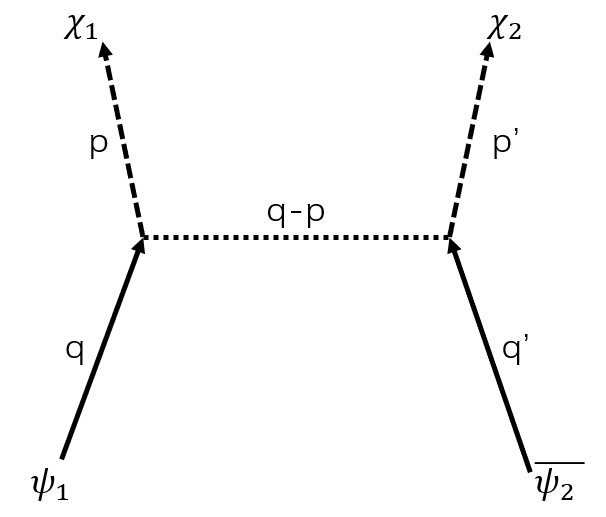}
    \caption{Feynman diagram for the direct scattering $\psi_1 {\bar\psi}_2 \to \chi_1 \chi_2$. The dotted line shows the virtual pair of $\phi$ particle and antiparticle.}
\end{figure}

We have
\begin{align}
\mathcal{M}^{(1)}= Z_\phi \frac{\beta_\phi^*}{\alpha_\phi^*}\int dx dt g_1(t) \int dx' dt' g_2(t') \int dk \bra{0,out} c_{p} c_{p'} \{\phi^\dagger\psi_1 \chi_1\}_{xt} \{\phi \psi_2^\dagger \chi_2\}_{x't'} d_q^\dagger f_{q'}^\dagger a_k^{out \dagger} b_{-k}^{out \dagger} \ket{0,out},
\label{M(1)}
\end{align}
where we allowed for the time dependence of the couplings $g_1$, $g_2$ and used curly brackets to group the fields depending on $x,t$ and on $x',t'$.
A straightforward calculation gives
\begin{align}
\mathcal{M}^{(1)}= Z_\phi \frac{\beta_\phi^*}{\alpha_\phi^*}\delta(p+p'-q-q') \int dt g_1(t)\psi_q^{out} (t)\phi_{q-p}^{out}(t)\chi_{-p}^*(t) \int dt' g_2(t')\psi_{-q'}^{out}(t')\phi_{p'-q'}^{out}(t') \chi_{-p'}^*(t'),
\label{M(1)}
\end{align}
{where we have used an expansion of the field operators $\psi$ and $\phi$ in out-vacuum modes.}

\subsection{Constant couplings}

We first consider the case of constant couplings, $g_i={\rm const}$.  Then in the center of mass frame of the two detector particles, $q+q'=0$, the two integrals in Eq.~(\ref{M(1)}) are equal to one another (apart from the constant factors $g_i$) and we obtain
\begin{align}
\mathcal{M}^{(1)}_{cm}= g_1 g_2 Z _\phi\frac{\beta_\phi^*}{\alpha_\phi^*}\delta(p+p') \mathcal{A}^2,
\label{M2detectors}
\end{align}
where
$\mathcal{A}$ is the same as in Eq.~(\ref{Asingle}) for single detector and is approximately given by (\ref{Aapprox}).  Hence we obtain
\begin{align}
\abs{\mathcal{M}^{(1)}_{cm}}= \delta(p+p')\abs{Z _\phi}\frac{\abs{\beta_\phi}}{{\abs{\alpha_\phi}}}\frac{\pi g_1 g_2}{2eE m_\chi^2 \omega_p}.
\label{Mcm}
\end{align}
and
\begin{align}
\label{cMcm}
    \bra{f}c^{\dagger}_{p(1)} c_{p(1)} \ket{f}=\delta(0) {\abs{Z _\phi}}^2{\abs{\frac{\beta_\phi}{\alpha_\phi}}}^2 \frac{\pi^2 g_1^2 g_2^2}{(2eE)^2 m_\chi^4 \omega_p^2}.
\end{align}

As before, the infinite factor $\delta(0)$ appears in the above equation because we used plane wave states normalized to a $\delta$-function.
Finitely normalized states could be obtained if instead we used periodic boundary conditions with a large periodicity scale $L$.  The momentum spectrum would then be discrete and the states would be normalized as
\begin{align}
\bra{q}\ket{q'}=\delta_{q,q'} (L/2\pi),
\end{align}
where $\delta_{q,q'}$ is the Kroneker delta.
Then the $\delta(0)$ gets replaced by $L/2\pi$ and the denominator of Eq.~(\ref{Nchi}) is replaced by
\begin{align}
\bra{q,{\bar q}'}\ket{q,{\bar q}'} = (L/2\pi)^2 .
\end{align} 
Hence the momentum distribution of $\chi_1$-particles becomes
\begin{align}\label{Nchip}
    \frac{dN_{\chi}}{dp}=  {\abs{Z _\phi}}^2{\abs{\frac{\beta_\phi}{\alpha_\phi}}}^2\frac{\pi^3 g^2_1g^2_2} {2e^2E^2m^4_{\chi}}\frac{1}{L\omega_{p}^2}.
\end{align}
It is clear from the symmetry that the distribution of $\chi_2$ particles in the center of mass frame is the same.

It is interesting to compare this distribution with that for a single detector particle, Eq.~(\ref{Nchipsingle}).
The dimensionless scattering amplitudes $g_1^2/eEm_\chi^2$  and $g_2^2/eEm_\chi^2$ in Eq.~(\ref{Nchip}) indicate that scatterings with both detector $\psi$-particles are involved.  The factor $\abs{\beta_\phi}^2$ accounts for the probability of Schwinger pair creation.  Note that there is only one factor of $\abs{\beta_\phi}^2$ in (\ref{Nchip}), indicating that both detector particles scatter on $\phi$-particles of the same pair.  The factor ${\abs{Z_\phi}}^2$ is the probability that no other pairs are produced.
The factor $L$ in the denominator can be understood as follows.  Our initial state of plane waves for the two detector particles corresponds to uniform fluxes of $\psi$ and ${\bar\psi}$ particles.  If $\psi$ hits a $\phi$-antiparticle, then in order for ${\bar\psi}$ to hit the particle of the same pair, its impact parameter has to be appropriately fine-tuned within a finite length $\Delta x$.  Then $dN_\chi/dp$ should be proportional to $\Delta x/L$.  (Eq.~(\ref{Nchip}) suggests that the impact parameter should be localized within $\Delta x \sim \omega_{p}^{-1}$.)  In other words, the factor $1/L$ can be interpreted as the density $n$ of detectors, and the number of $\chi$-particles in the final state should be proportional to this density.


\subsection{Time dependent couplings}

We now consider time dependent couplings, $g_i(t)=g_i f(t-\tau_i)$ with $f(t)$ from Eq.~(\ref{ft}), so each detector is turned on only for a finite period of time $\sim T$ around $t=\tau_i$ ($i=1,2$).  
Then we can rewrite Eq.~(\ref{M(1)}) as
\begin{align}
\mathcal{M}^{(1)}= g_1 g_2 Z_\phi \frac{\beta_\phi^*}{\alpha_\phi^*}\delta(p+p'-q-q') \mathcal{A}_1\mathcal{A}_2,
\label{M(1)2}
\end{align}
where 
\begin{align}
\mathcal{A}_1=\int dt f(t-\tau_1)\psi_q^{out} (t)\phi_{q-p}^{out}(t)\chi_{-p}^*(t),
\label{Atau1}
\end{align}
\begin{align}
\mathcal{A}_2=\int dt f(t-\tau_2)\psi_{-q'}^{out} (t)\phi_{p'-q'}^{out}(t)\chi_{-p'}^*(t).
\end{align} 
Now, it follows from the expression for the mode functions (\ref{phikout}) that $\phi_k^{out}(t)=\phi_{k+eE\tau}^{out}(t-\tau)$ and $\psi_q^{out}(t)=\psi_{q+eE\tau}^{out}(t-\tau)$.  Substituting this in Eq.~(\ref{Atau1}) and changing the integration variable $t\to t+\tau_1$ and then $t\to -t$, we obtain 
\begin{align}
\mathcal{A}_1=e^{i\omega_p \tau_1}\int dt f(t)\psi_{{\tilde q}_1}^{out} (t)\phi_{{\tilde q}_1-p}^{out}(t)\chi_{-p}^*(t) 
=\int dt f(t)\psi_{-{\tilde q}_1}^{in*} (t)\phi_{p-{\tilde q}_1}^{in*}(t)\chi_{-p}(t),
\label{Atau11}
\end{align} 
where ${\tilde q}_1=q+eE\tau_1=q_{phys}(\tau_1)$ and we have used the symmetry $f(-t)=f(t)$ and dropped a constant phase factor in the second step.  Similarly, we find
\begin{align}
\mathcal{A}_2=\int dt f(t)\psi_{{{\tilde q}'}_2}^{in*} (t)\phi_{{\tilde q}'_2-p'}^{in*}(t)\chi_{-p'}(t),
\label{Atau22}
\end{align} 
where ${\tilde q}'_2=q'-eE\tau_2=q'_{phys}(\tau_2)$.
${\tilde q}_1$ is the physical momentum of the $\psi$-detector and ${\tilde q}'_2$ is the physical momentum of the ${\bar\psi}$ detector at the times of their respective collisions.
	
The integral in the last step of Eq.~(\ref{Atau11}) is the same as in Eq.~(\ref{A-p-q}) for a single $\psi$ detector with $q$ replaced by ${\tilde q}_1$.  In fact the two integrals are identical, since Eq.~(\ref{A-p-q}) was derived for $\tau_1=0$, in which case ${\tilde q}_1=q$.
Similarly, it is easily verified that Eq.~(\ref{Atau22}) is the same as we would get for a single ${\bar\psi}$ detector.
Eq.~(\ref{M(1)2}) tells us that apart from the overall momentum conservation the measurements of the two detectors are independent from one another.  In particular, it follows from the results of Refs.~\cite{Garriga1,Garriga2} that the $\psi$ detector will find ${\bar\phi}$ particles moving mostly to the left and the ${\bar\psi}$ detector will find $\phi$ particles moving mostly to the right (in the respective rest framers of the detectors).  This is perhaps not surprising, since semiclassically the two collision events 
are spacelike separated.  

To simplify further discussion, we shall assume that the detector particles have equal and opposite momenta, $q+q'=0$, and that the detectors are turned on at the same time, $\tau_1=\tau_2$.  
Then we have $p'=-p$, ${\tilde q}'_2=-{\tilde q}_1$, and $\mathcal{A}_2=\mathcal{A}_1\equiv \mathcal{A}$. 
The physical momenta of the detected ${\bar\phi}$ and $\phi$ particles are respectively $k_{phys}=p-q_{phys}$
and ${k'}_{phys}=p'-{q'}_{phys}=-k_{phys}$.  

An interesting situation arises when the physical momentum of the $\psi$ detector is large and positive and the physical momentum of the ${\bar\psi}$ detector is large and negative, so that $k_{phys}>0$ and $k'_{phys}<0$.  We shall further assume that the magnitudes of both physical momenta satisfy the conditions (\ref{condition4}), (\ref{condition3}). 
Then the amplitude $\mathcal{A}$ is well approximated by Eq.~(\ref{A2}):
\begin{align}
    \mathcal{A}   \approx \frac{\sqrt{\pi} T}{\sqrt{8{\tilde q} (p-{\tilde q})\omega_{p}}}\exp\left[-\frac{T^2}{4}(\epsilon_{{\tilde q}_i}+\xi_{{\tilde q}-p} -\omega_{p})^2\right],
\end{align}
As we discussed in Sec. III.B, this amplitude as functions of $p$ has a peak at some value $p_*$, which depends on ${\tilde q}$.  If $p$ is not far from $p_*$, we can use the approximate form (\ref{Aapprox3}) for $\mathcal{A}$:  
\begin{align}
    \mathcal{A} \approx \sqrt{\frac{\pi}{2}}\frac{m_\chi T}{2m_\phi {\tilde q}^{3/2}}\exp\left[-\frac{m_\chi^8 T^2}{16m_\phi^4 {\tilde q}^4} \left(p-{\tilde q}-\frac{m_\phi^2}{m_\chi^2}{\tilde q} \right)^2\right].
\end{align}
The total transition amplitude is then
\begin{align}
\mathcal{M}^{(1)}_{cm} \approx Z_\phi {\frac{\beta_\phi^*}{\alpha_\phi^*}}\delta(p+p')  \frac{\pi g_1 g_2 m^2_\chi T^2}{8m_\phi^2 q^{3}}\exp\left[-\frac{m_\chi^8 T^2}{8m_\phi^4 {\tilde q}^4} \left(p-{\tilde q}-\frac{m_\phi^2}{m_\chi^2}{\tilde q} \right)^2\right].
\label{Mapprox}
\end{align}
and the distribution of $\chi_1$ particles is given by
\begin{align}
\frac{dN_\chi}{dp} \approx \left|Z_\phi \frac{\beta_\phi^*}{\alpha_\phi^*}\right|^2 
\frac{\pi^3 g_1^2 g_2^2 m^4_\chi T^4}{32 m_\phi^4 {\tilde q}^{6}L}\exp\left[-\frac{m_\chi^8 T^2}{4m_\phi^4 {\tilde q}^4} \left(p-{\tilde q}-\frac{m_\phi^2}{m_\chi^2}{\tilde q} \right)^2\right].
\label{dNdchi2}
\end{align}
Eqs.~(\ref{Mapprox}) and (\ref{dNdchi2}) indicate that for sufficiently large $q_{phys}$ the particles of the pair can be observed moving towards one another with arbitrarily large momenta $k_{phys}=-k'_{phys}>0$. The corresponding amplitude is not exponentially suppressed.  This conclusion could be anticipated: it follows from our results for a single detector in Sec.~III.B and from the fact that measurements of the two detectors are independent.

In a more general case, when the conditions (\ref{condition4}),  (\ref{condition3}) are not satisfied, {we note that to the extent that energy is conserved during the collision,} we can use the classical Eq.~(\ref{kq}) to find a general condition for the particles of the pair to move towards one another.  This will happen when $k_{phys}>0$ for both signs in Eq.~(\ref{kq}), that is, when
\begin{align}
v\equiv \frac{q_{phys}}{\sqrt{q_{phys}^2 +m_\psi^2}} > \sqrt{1-\frac{4m_\phi^2 m_\psi^2}{M^4}}.
\label{cond}
\end{align}
The quantity $v$ is the velocity of the $\psi$ detector; the corresponding condition for the Lorentz factor, $\gamma = (1-v^2)^{-1/2}$, is
\begin{align}
\gamma >\frac{M^2}{2m_\phi m_\psi}.
\label{gammacond}
\end{align}

We finally comment on what happens if $\psi$ and ${\bar\psi}$ detectors are turned on at different times, $\tau_1\neq\tau_2$.  In this case the amplitude $\mathcal{M}^{(1)}_{cm}$ is proportional to 
\begin{align}
\exp\left[-\frac{m_\chi^8 T^2}{16m_\phi^4 {\tilde q}_1^4} \left(p-{\tilde q}_1-\frac{m_\phi^2}{m_\chi^2}{\tilde q}_1 \right)^2\right] \exp\left[-\frac{m_\chi^8 T^2}{16m_\phi^4 {\tilde q}_2^4} \left(p-{\tilde q}_2-\frac{m_\phi^2}{m_\chi^2}{\tilde q}_2 \right)^2\right],
\end{align}
where ${\tilde q}_i=q+eE\tau_i$.  If $\tau_2$ is significantly different from $\tau_1$, the two peaks do not overlap and the amplitude is exponentially suppressed (compared to that for $\tau_2=\tau_1$).  The reason for this is that for a given value of $p'=-p$, the classical kinematics allows only a single value of ${\tilde q}'=-{\tilde q}$, which corresponds to the two scatterings occurring at the same time.

\section{Conclusions}

We applied the detector model similar to that used in Refs.~\cite{Garriga1,Garriga2} to study pair production of charged particles in a constant electric field in $(1+1)$ dimensions.  We considered the situation where two different detectors moving relative to one another are used to observe the particle and antiparticle of the same pair.  Our conclusion is that each detector observes nucleation to occur in its rest frame, regardless of the motion of the other detector.   But if the relative velocity of the two detectors is sufficiently high, the particle and antiparticle can be observed moving towards one another.  This is a surprising result.  

In our model the $\phi$ and ${\bar\phi}$ particles are destroyed at the moment of detection.  But in the semiclassical regime it may be possible to construct detectors capable of observing the particles with relatively small interference, so that they are not destroyed and their momentum is not significantly changed during their interaction with the detector.  Then, once a particle is detected, we would be able to use successive measurements to observe the rest of particle's trajectory, as we would in a bubble chamber.  
An observer watching both particles may then first see them moving towards one another, stop momentarily at separation $2r_0$ and then move apart.  This goes against the conventional wisdom, since it is usually assumed that in any reference frame the part of the history of the pair where the particle and antiparticle move towards one another is unobservable.

One might expect similar conclusions to apply to the case of bubble nucleation in $(3+1)$ dimensions. If different parts
of a bubble are observed by differently moving observers, each observer would then see the bubble wall approaching from the direction of the bubble center.  However, when measurements made by different observers are combined, the resulting nucleation hypersurface would generally be rather irregular.  If the relative velocities of the observers are sufficiently high, the bubble worldsheet above this hypersurface may include a large contracting region, which is usually discarded.  

The appearance of contracting pairs (that is, pairs in which $\phi$ and ${\bar\phi}$ move towards one another) raises some interesting questions.\footnote{We are grateful to Jaume Garriga for an illuminating discussion of these questions.}   We found that such pairs can be observed with the aid of two detectors approaching one another with a large Lorentz factor.  The question is: are such pairs still there when nobody is looking?  The answer to this is "No", if "being there" is understood in the (semi)classical sense.  The semiclassical trajectory of a charged particle generically has two encounters with an inertial observer, one before and one after the turning point. Before the turning point, the particle would move to the right relative to the observer, while after the turning point it would move to the left. However we find that our ${\bar\phi}$ particles move predominantly to the left at the moment of detection, regardless of the state of motion of the detector. This suggests that the semiclassical trajectory was not there before detection. 

Once ${\bar\phi}$ is detected, it may closely follow its classical trajectory (assuming a non-invasive detector that we alluded to above.) After that, other detectors may see it moving to the left as well as to the right, depending on the state of detector motion. However, in the first detection ${\bar\phi}$ is (almost) always observed moving to the left -- which  corresponds to the expanding part of ${\bar\phi}$ trajectory in the frame of the detector.

A related question is: could it be that the creation of a contracting pair is somehow triggered by the rapidly moving detectors?  Once again, the answer is "No".   We used the representation of the in-vacuum in terms of out-states, and the only term that contributed to the detection amplitude is the one proportional to $Z_\phi \beta_\phi$, that is, to the amplitude of spontaneous creation of a pair.  We thus see no indication of a stimulated pair production.


Given the unexpected nature of our results, one should keep in mind possible caveats. A potential problem with our approach is that, as noted in Ref.~\cite{Garriga1}, the in-vacuum state that we assumed here has some unphysical properties.  The singularity structure of the two-point function in this state does not have the Hadamard form, and as a result the expectation values of physical observables cannot be regulated in a Lorentz invariant way.  An important special case is that of the electric current.  One expects that the created pairs moving in the electric field will develop a nonzero current, which will break the Lorentz invariance.  And indeed one can show that all physical (Hadamard) states of charged particles in a constant electric field are not Lorentz invariant.  

Physically, this issue is related to the fact that a metastable vacuum could not have existed for an infinite time.
In a more physical approach the electric field would have to be time dependent with $E(t\to -\infty)\to 0$, so the initial state of the $\phi$ could be chosen as the standard vacuum state.  The two-point function is then Hadamard at $t\to -\infty$ and is guaranteed to remain Hadamard at later times.  It would be interesting to repeat our analysis (and that of Refs.~\cite{Garriga1,Garriga2}) in this setting.  The time dependence of the electric field introduces a preferred frame and explicitly breaks the Lorentz invariance.  The question is to what extent this Lorentz violation influences the results.  We leave this as a problem for future research.

\bigskip

{\bf Acknowledgements}

\bigskip

We are grateful to Larry Ford, Jaume Garriga, Sugumi Kanno and Takahiro Tanaka for very useful discussions.  This work was supported in part by grant PHY-1820872 from the National Science Foundation.

\appendix
\section{}

To show the result in (\ref{Aapprox}), we follow the calculation in Appendix B of \cite{Garriga2}. The 4-point amplitude in \cite{Garriga2} is defined as
\begin{align}
    \mathcal{A}_2(q,k,\Tilde{p};\Tilde{p}')=g\int^{\infty}_{-\infty}dt \phi^{in*}_k(t)\psi^{in*}_q(t)\chi_{1,\Tilde{p}'}(t)\chi^*_{2,\Tilde{p}}(t).
\end{align}
This amplitude can be reduced to the amplitude $\mathcal{A}$ that we adopt here by dropping the factor $g$ and setting $\chi^*_{2,\Tilde{p}}(t)\to1$. To achieve that, $\omega_{\Tilde{p}}$ is set to be zero in the exponential and $\Tilde{p}'$ is replaced by $p$ in (B.1) in \cite{Garriga2}. Meanwhile, the amplitude is also multiplied by an additional $\sqrt{2\omega_{\Tilde{p}}}$. Hence, we obtain
\begin{align}
    \abs{\mathcal{A}}^2=\frac{\pi e^{-\frac{\pi}{2}\lambda_{\phi}}e^{-\pi\lambda_{\psi}}}{2eE m^2_{\chi}\omega_p}e^{-\pi\sigma_-}\abs{W_{-i\sigma_+,i\sigma_-}\left(e^{i\frac{3}{2}\pi}\frac{m^2_{\chi}}{2eE}\right)}^2,
\end{align}
where $W_{-i\sigma_+,i\sigma_+}$ is the Whittaker function and $\sigma_{\pm}\equiv(\lambda_{\phi}\pm\lambda_{\psi})/4$. Due to the mass hierarchy, an asymptotic representation can be applied: $W_{\lambda,\mu}(z)\approx e^{z/2}z^{\lambda}$, when $\abs{z}^2\gg\abs{\mu^2-\lambda^2}$. Thus, we have
\begin{align}
    \abs{\mathcal{A}}^2=\frac{\pi}{2eE m^2_{\chi}\omega_p}.
\end{align}

\bibliographystyle{unsrt}
\bibliography{mybib}

\end{document}